\documentclass[twocolumn,showpacs,preprintnumbers,amsmath,amssymb,floatfix]{revtex4}
\usepackage{graphicx}
\usepackage{dcolumn}
\usepackage{bm}
\usepackage{citesort}
\usepackage{array}

\begin{document}

\preprint{Version 2}

\title{Thermodynamics of a Spin-1/2 Chain Coupled to Einstein Phonons}

\author{Alexander B\"uhler}
 \email{ab@thp.uni-koeln.de}
 \affiliation{Institut f\"ur Theoretische Physik, %
   Universit\"at zu K\"oln, %
   Z\"ulpicher Stra{\ss}e 77, %
   50937 K\"oln, %
   Germany}
\author{Jaan Oitmaa}
 \email{otja@phys.unsw.edu.au}
 \affiliation{School of Physics, %
   The University of New South Wales, %
   Sydney, %
   NSW 2052, %
   Australia}
\author{G\"otz S. Uhrig}
 \homepage{http://www.thp.uni-koeln.de/~gu}
 \email{gu@thp.uni-koeln.de}
 \affiliation{Institut f\"ur Theoretische Physik, %
   Universit\"at zu K\"oln, %
   Z\"ulpicher Stra{\ss}e 77, %
   50937 K\"oln, %
   Germany}

\date{\today}

\begin{abstract}
A high order series expansion is employed to study the
thermodynamical properties of a $S=1/2$ chain coupled to dispersionless
phonons. The results are obtained without truncating the phonon
subspace since the series expansion is performed formally
in the overall exchange coupling $J$. The results are used to
investigate various parameter regimes, e.g.\ the adiabatic and
antiadiabatic limit as well as the intermediate regime which is 
difficult to investigate by other methods.
We find that dynamic phonon effects become manifest when more than
one thermodynamic quantity is analyzed.
\end{abstract}



\maketitle

\section{\label{sec:intro} Introduction}
In solid state physics, all electronic degrees of freedom like charge
or spin are coupled to 
vibrations. Mostly, however, such a coupling does not influence the
system's properties in a decisive way. This is different if the
electronic degrees of freedom are essentially one dimensional.
Then the phenomenon of a Peierls transition occurs: the system
breaks translational invariance spontaneously by forming dimers
\cite{frohl54,peier55,pytte74b,su79,cross79,bray83}.
The interest in a model of quantum phonons coupled to one-dimensional spin 
degrees of freedom in particular has been rekindled by  the discovery of the
first inorganic spin-Peierls substance CuGeO$_{\text{3}}$
\cite{hase93a}. Single crystals of high quality made
investigations possible that were not possible for the long known
organic spin-Peierls substances \cite{bray83}.
 
Besides the spin-Peierls phenomenon a very strong  coupling of spins and
phonons can influence the quantitative physics of Mott insulators significantly
if the superexchange coupling is small due to geometrical reasons.
Examples are  a 90$^\circ$ angle in the exchange path or a complicated
superexchange process via large ligand groups. 
In these cases,  a small change of the geometry implies a certain change
of the coupling which is  very large \emph{relative} to the unchanged coupling.
Hence the influence of phonons is much larger than usual. Examples for this
mechanism besides CuGeO$_3$\cite{brade96a,geert96} 
are  SrCu$_2$(BO$_3$)$_2$ \cite{kagey99a,miyah00b} or
(VO)$_2$P$_2$O$_7$ \cite{garre97a,grove00,uhrig01a}.

For the above reasons, it is of significant interest to provide
reliable theoretical predictions for spin systems coupled to phonons.
It is also clear that acoustic phonons will not have a significant
influence because they alter the exchange pathes relatively weakly.
Thus we focus on optical phonons which have a strong impact on the local
geometry so that they can influence the exchange coupling significantly.
For simplicity, we will consider dispersionless Einstein phonons.
The aim of the present work is to compute the two fundamental
thermodynamic quantities of spin systems, namely the magnetic
susceptibility and the specific heat, and to compare the results
in presence of a spin-phonon coupling to the results of static
spin models. This comparison serves as a  guideline to experimental 
analyses which attempt to identify the signatures of spin-phonon couplings.

The generic spin-phonon model introduced in the following section cannot be
solved analytically. To the authors' knowledge, there are no analytical
exact methods to treat extended systems of coupled spins and phonons
if all energy scales shall be considered.
Many approximate methods have been applied such as density-matrix
renormalization \cite{bursi99a}, continuous unitary transformations
\cite{uhrig98b,raas01a,raas02a}, exact diagonalization \cite{weiss99b},
linked cluster expansion \cite{trebs01}, renormalization group
\cite{sun00} and quantum Monte Carlo (QMC)
\cite{raas02a,sandv97a,kuhne99b,sandv99a,aits02,aits03}.

Two limits can be analyzed in more detail. In the adiabatic limit 
$\omega \ll J$ the spin system is assumed to be `fast' compared to the `slow' 
phonon system. Using approaches analogous to the ones applied by Pytte 
\cite{pytte74b} and to the more detailed one by Cross and Fisher 
\cite{cross79} the model in Eq.~(\ref{eqn:hamilton}) can be mapped to a 
statically dimerized model. 

The antiadiabatic limit $\omega \gg J$ can be 
handled by an appropriate mapping of the starting Hamiltonian to a frustrated 
spin model. Thereby, interactions of larger range are induced and the 
phonon frequency is renormalized
\cite{uhrig98b,weiss99b,bursi99a,buhle99,raas01a,raas02a}. Above a
critical frustration, e.g.\ a certain next-nearest neighbor interaction, the
system becomes gapped. This regime is reached for large values of
the spin-phonon coupling. For small values of the spin-phonon coupling
the system remains gapless. Note that we are dealing with a
purely one-dimensional problem so that quantum fluctuations may 
prevent spontaneous symmetry breaking.

Concerning thermodynamic properties it 
could be shown that the magnetic susceptibility can be fitted
by a frustrated spin model with temperature independent couplings. But
this approach fails for increasing values $J/\omega$
\cite{kuhne99b,raas02a}.

Investigations of the regime between the adiabatic and the antiadiabatic
limit with $\omega \approx J$ are difficult. So far, a renormalization group 
analysis \cite{sun00} and quantum Monte-Carlo simulations 
\cite{kuhne99b,raas02a} have been done.

In the present paper, a spin-phonon model is studied using a 
linked-cluster expansion to derive thermodynamical properties like the 
specific heat $C$ and the susceptibility $\chi$. Zero temperature
properties were determined previously \cite{trebs01}.
The paper is organized as follows. Sect.~\ref{sec:method} presents the
model, some of its known properties and the basic elements of the
method employed. In Sect.~\ref{sec:results} the results
are analyzed. They are summarized in Sect.~\ref{sec:conclusion}
where also open issues are identified.

\section{\label{sec:method} Model and Method}
\begin{figure}[htbp]
  \begin{center}
    \leavevmode
    \includegraphics[width=\columnwidth]{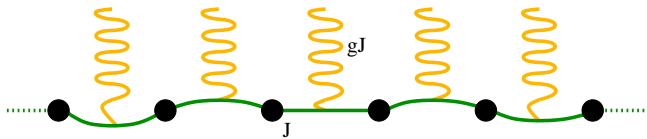}
    \caption{Schematic picture of the spin-phonon model as defined by
    Eq.~(\ref{eqn:hamilton}).} 
    \label{fig:model}
  \end{center}
\end{figure}

The isotropic spin-1/2 Heisenberg chain is extended by the coupling
to local, dispersionless phononic degrees of freedom. 
The Hamilton operator reads
\begin{subequations}
  \label{eqn:hamilton}
  \begin{eqnarray}
    H &=& J\sum_i(1 + g(b_i^{\dagger} +
     b_i^{\phantom{\dagger}}))\mathbf{S}_i\mathbf{S}_{i+1} +
     \omega\sum_i b_i^{\dagger}b_i^{\phantom{\dagger}}\\ 
     &=&   H_{\text{SB}} + H_{\text{B}} \ .
  \end{eqnarray}
\end{subequations}
The magnetic exchange coupling is denoted by $J$, the coupling
between the phononic subsystem and the magnetic subsystem is given by
$gJ$, and the energy of the phonons is $\omega$. The
abbreviations $H_{\text{B}}$ and $H_{\text{SB}}$ are used in the
following. The Hamiltonian in 
Eq.~(\ref{eqn:hamilton}) represents the so-called bond-coupling
model depicted schematically in Fig.~(\ref{fig:model}). The 
phonons can be viewed to sit between the spin sites. They influence
only one bond. Such a coupling can be motivated microscopically
\cite{werne99} for CuGeO$_3$. But the main reason to choose the coupling
as in Eq.\ (\ref{eqn:hamilton}) for our study is its simplicity.

In the antiadiabatic limit $J/\omega \rightarrow 0$, the critical 
spin-phonon coupling $g_c$ for a phase transition from a gapless to a gapped 
phase is given by $g_c/\omega \approx 0.4682$ using the flow equation approach
 \cite{raas02a}. Assuming a dimerized phase, the model was investigated at
$T=0$ \cite{trebs01} by a linked-cluster expansion which avoided any
truncation of the phononic Hilbert space. But the starting point
was the symmetry-broken dimerized phase at zero temperature.
In the present article, we emphasize the thermodynamical aspects of the 
model without broken symmetry. 
A series expansion about the limit of vanishing $J/T$ is performed. The
phononic subspace is treated exactly. No cutoff in the
the phonon subspace is necessary. The resulting quantities are given
as truncated series in $J/T$ with full dependence on the remaining
parameters such as $\omega/T$.

First we calculate the partition function $Z$ of
the spin-phonon system. Then, quantities like the free energy,
the specific heat or the susceptibility can be derived easily. 
A traditional high temperature series expansion is not
possible because the expansion in the inverse temperature would lead to
divergences in the phononic degrees of freedom. The limit of infinite
temperature is not a well-defined starting point for bosons since
the phonon occupation number diverges. For this reason, we
choose to perform the formal expansion in the exchange coupling $J$.
In this way, the phonon subspace is treated exactly for each given 
temperature. No truncation is necessary and the full phonon dynamics is taken 
into account. To our knowledge, this is the first approach of a cluster
expansion about the limit $J=0$ at finite temperatures which avoids
approximations in the phonon subspace. 

We stress that the formal series in $J$ coincides with the series
for the magnetic subsystem in the inverse temperature if we set the
spin-phonon coupling to zero: $g=0$. So it is not surprising that
the obtained series bears many similarities to a high-temperature
series. It is most reliable at high temperatures. The limit to
vanishing temperature is difficult to describe.

The Hamilton operator from Eq.~(\ref{eqn:hamilton}) is split into its
diagonal part $H_0=H_{\text{B}}$ and a perturbation $V$ with
\begin{subequations}
  \label{eqn:hamiltoninter}
  \begin{eqnarray}
    H &=&  H_0+ JV = H_{\text{B}} + H_{\text{SB}}\\ 
    &=& \underbrace{\omega\sum_i
    b_i^{\dagger}b_i^{\phantom{\dagger}}}_{H_0} 
      +J \underbrace{\sum_i(1 + g(b_i^{\dagger} +
     b_i^{\phantom{\dagger}}))\mathbf{S}_i\mathbf{S}_{i+1}}_V       
  \end{eqnarray}
\end{subequations}
The diagonal part $H_0$ is trivially solvable; it describes free
dispersionless (Einstein) phonons. The perturbation $V$ includes the
isolated magnetic part and the spin-phonon interaction. The standard
way to treat such a problem is to change to the interaction
representation where the off-diagonal perturbation governs the
non-trivial part of the dynamics of the system. 
In this framework the partition function is given as an infinite series 
in the expansion parameter $J$ by
\begin{subequations}
    \label{eqn:partition}
    \begin{eqnarray}
      Z \!\!\!&=&\!\! \text{tr}\left\{ e^{-\beta H}\right\}\\
      &=&\!\!\! Z_0\Big(\!1\!+\!\!
      \sum_{n=1}^{\infty}\!\left(-J\right)^n \!\!
      \int\limits_0^{\beta}\!\!d\!\tau_1\cdots \!\!\!
      \int\limits_0^{\tau_{n-1}}\!\!\!\!d\!\tau_n \langle\tilde{V}(\tau_1)
      \cdots \tilde{V}(\tau_n) \rangle \Big) \hspace*{4mm}
  \end{eqnarray}
\end{subequations}
where the following abbreviations are used. The unperturbed part
$H_0$ in (\ref{eqn:hamiltoninter}) leads to the
 contribution $Z_0$ of the partition function 
\begin{equation}
  \label{eqn:partition0}
  Z_0 = \text{tr}\left\{ e^{-\beta H_0}\right\} = 2^N \left\{ \prod_i
  \left( \sum_{n_i=0}^\infty
  e^{-\beta\omega n_i}\right)\right\} = 2^Nz_0^N\
\end{equation}
with $z_0 = 1/(1-e^{-\beta \omega})$ and the phonon occupation number
$n_i = b_i^{\dagger}b_i^{\phantom{\dagger}}$.
The system size is denoted by $N$. The perturbation $V$ given in the
interaction representation as $\tilde{V}$ reads
\begin{subequations}
  \label{eqn:Vtilde}
  \begin{eqnarray}    
    \tilde{V}(\tau) &=& e^{\tau H_0}V e^{-\tau H_0} \phantom{\sum_i}
      \\ &=& \sum_i\mathbf{S}_i\mathbf{S}_{i+1} \left( 1+ g
      e^{\tau H_0}\left( b_i^{\dagger} +
      b_i^{\phantom{\dagger}}\right) e^{-\tau H_0} \right) \\ &=&
      \sum_i\mathbf{S}_i\mathbf{S}_{i+1} \left( 1+ g\left(
      b_i^{\dagger} e^{\omega\tau}+ b_i^{\phantom{\dagger}}
      e^{-\omega\tau} \right) \right)\ .
  \end{eqnarray}
\end{subequations}
The angular brackets in Eq.~(\ref{eqn:partition}) are an
abbreviated notation for
\begin{equation}
  \label{eqn:Vbrackets}
  \langle\tilde{V}(\tau_1)\cdots\tilde{V}(\tau_n) \rangle =
  \frac{1}{Z_0} \text{tr}\left\{e^{-\beta H_0}
  \tilde{V}(\tau_1)\cdots\tilde{V}(\tau_n) \right\}\ .
\end{equation}
As can be seen from the above equations the calculations for the
partition function $Z$ of the magnetic system and of the phononic
system factorize. In each order of expansion in $J$ the contribution
from the spin system can be evaluated separately from the phononic
contributions.

Calculating the partition function in Eq.~(\ref{eqn:partition})
requires repeated integrations over functions of the type 
\begin{equation}
  \label{eqn:integrand}
  I(k,l;x_n) = x_n^ke^{l x_n}\ \text{ with }\ k\in \mathbb{N}_0,\
  l\in\mathbb{Z},\ x_n\in\mathbb{R}\ .
\end{equation}
The resulting integrals can be solved exactly with
\begin{subequations}
  \label{eqn:integral}
  \begin{eqnarray}
    l\neq 0: && \int\limits_0^{x_{n-1}}\!d\!x_n x_n^ke^{lx_n}  =
    k!\left(- \frac{1}{l}\right)^{k+1} \nonumber \\
    && + \sum_{i=0}^{k}\left(
    -1\right)^i \frac{1}{l^{i+1}} \frac{k!}{(k-i)!} x_{n-1}^{k-i}
    e^{lx_{n-1}}   \\  
    l = 0: && \int\limits_0^{x_{n-1}}\!d\!x_n x_n^k =
    \frac{1}{k+1}x_{n-1}^{k+1} \ .
  \end{eqnarray}
\end{subequations}
These equations allow an iterative evaluation of the multiple
integrals entering the partition function $Z$.

A useful check of the calculations is the limit $g=0$. This special
case yields 
\begin{equation}
  \label{eqn:geq0}
  Z_{g=0} = Z_{\text{isol. phonons}}Z_{\text{isol. spins}} =
  z_0^NZ_{\text{isol. spins}}\ .
\end{equation}

Due to the one-dimensionality of the system under study a
simple cluster algorithm will be used (for an instructive review see
Ref.~\cite{domb74}). Therein not only the connected clusters are 
calculated but also the disconnected clusters. The unnecessary 
calculations of the disconnected cluster are not very costly and
we can save the book-keeping overhead which would be required otherwise.
The problem of  subtracting subclusters occurring in the linked cluster 
expansion algorithm is replaced by the evaluation of the lattice constants 
for a given cluster.

\section{\label{sec:results} Results} 
Here the results for the specific heat $C$ and for the
susceptibility $\chi$ are presented in separate subsections,
Sects.~\ref{ssec:heat} and \ref{ssec:susc}. A third subsection
Sect.~\ref{ssec:fit} is dedicated to the comparison of the results from static 
spin models and those from the spin-phonon model.

The bare truncated series provides a first impression of the behavior of the
considered quantities  for various sets of parameters. But for
quantitative predictions the truncated series are not sufficient as 
will be seen in the following. Extrapolation techniques are necessary
to improve the representation of the results for larger values of $J$.
It turns out that the description of higher temperatures is easily
possible whereas the extrapolation becomes ambiguous at low temperatures.
We attribute this behavior to the fact that the phononic and the
magnetic subsystem behave at higher temperatures more and more 
independently. Then our series is essentially a high-temperature 
expansion for the magnetic system which is known to work well
for higher temperatures.
For low temperatures, however, possible long-range effects set it
which elude our approach.

We benchmark our results relative to QMC data for selected sets of 
parameters \cite{aits02}.

The results for the spin-phonon model are compared to those of pure spin
systems. As a reference the exact result of the isotropic Heisenberg
model \cite{klump93b,egger94,klump98a} is depicted in the figures for
the susceptibility. The specific heat is compared to the specific heat
of free phonons plus the exactly known result for $C(T)$ for the Heisenberg
model \cite{klump98a}.

The coefficients of the series expansion results are available upon
request.

\subsection{\label{ssec:heat} Specific Heat}
A detailed study of the magnetic properties of the system under
consideration also includes the investigation of the specific
heat. Besides the magnetic susceptibility the specific heat is an
observable which can be experimentally easily measured and
theoretically easily calculated. A direct comparison between
theory and experiment is often hindered by the fact that the phononic
degrees of freedom dominate the specific heat. Our model 
(\ref{eqn:hamilton}) takes the influence of a strongly 
coupled optical phonon into account. The additional contribution of 
acoustic phonons is beyond the scope of the present investigation. 
We assume that the contribution of the acoustic phonons is indeed 
additive so that it can be accounted for if the lattice vibrations are 
known, e.g.\ from a dynamic lattice model.

The specific heat $C(T)$ is obtained from the results for the free
energy per site $f$ obtained from the partition function $Z$ using 
standard relations from statistical physics 
\begin{equation}
  \label{eqn_phonon:freeE}
  f = F/N= -\frac{1}{\beta}\frac{1}{N}\text{ln}Z\ .
\end{equation}
The free energy could be computed to order 13 in $J$. To this end, the
contribution of 214 connected and of 470 disconnected clusters had to
be evaluated. To illustrate the result the first orders of the free energy
series are given
\begin{eqnarray}
  \label{eqn:freeseries}
    && -\beta f = \frac{1}{N}\text{ln}(Z) = \text{ln}z_0 
    + J^2\left( \frac{3}{32}\beta^2 +
    \frac{3}{16}\frac{g^2\beta}{\omega} 
    \right) \nonumber \\
  &&\quad + J^3\left(\frac{1}{64}\beta^3   
    + \frac{3}{32}\frac{g^2\beta^2}{\omega}
    \right) \nonumber \\ 
  &&\quad + J^4 \frac{1}{256}\Big(
    -\frac{5}{4}{\beta}^{4}+6\frac{g^2\beta^3}{\omega} 
    \nonumber \\
  &&\quad + \left(\left( 24 z_0^2- 24
      {z_0}+ 6 \right)\frac{g^4}{\omega^2} +
    \left(-48{z_0} +24\right) \frac{g^2}{\omega^2}
    \right)\beta^2 
    \nonumber \\
  &&\quad + \left( \left(
      12 - 24 {z_0} \right)
    \frac{g^4}{\omega^3} +48\frac{g^2}{\omega^3} \right) \beta \Big) 
    + \mathcal{O}(J^5)\ .
\end{eqnarray}

\begin{figure}[htbp]
  \begin{center}
    \leavevmode
    \includegraphics[width=\columnwidth]{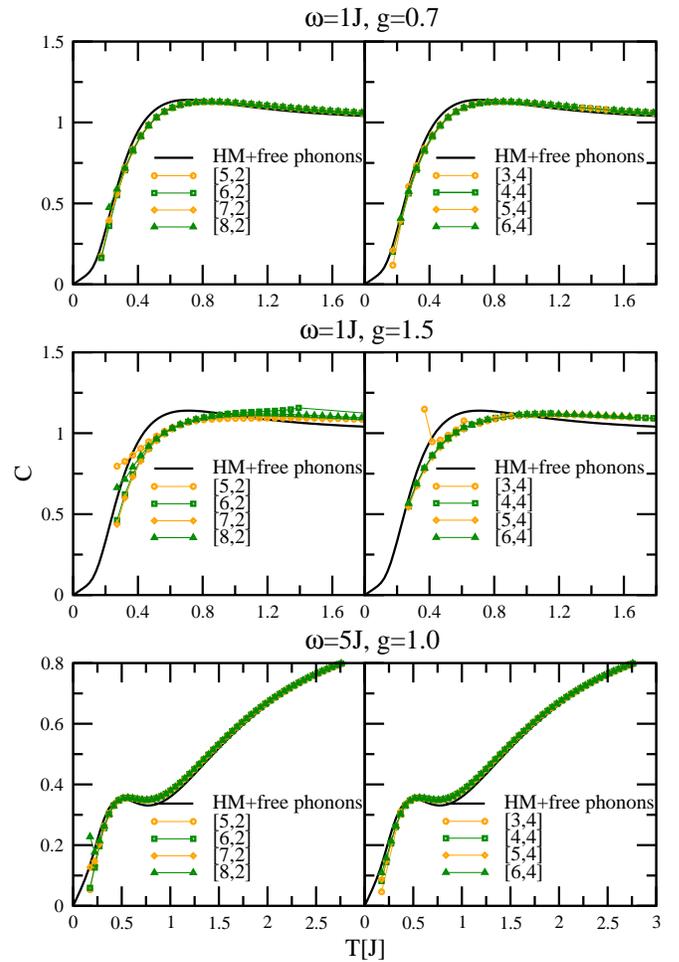} 
    \caption{Dlog-Pad\'e extrapolations of the specific heat: various
      extrapolations are compared for three different sets of
      parameters.} 
    \label{fig:spec.Dlog}
  \end{center}
\end{figure}

To obtain a detailed insight in the behavior of the specific heat
Dlog-Pad\'e extrapolations are used. The truncated series alone are
not appropriate for temperatures below $T < J$ (not shown). 
Fig.~\ref{fig:spec.Dlog} depicts the extrapolation results
of the specific heat compared to a superposition of the free phonon
part of the specific heat given by
\begin{equation}
  \label{eqn:Cfreephonons}
  C_B = \left(\beta\omega\right)^2\frac{e^{-\beta\omega}}{\left(
  1-e^{-\beta\omega}\right)^2}
\end{equation}
and the exactly known result $C_S$ for the isotropic Heisenberg
model. Various parameter sets are shown. Simple Dlog-Pad\'e
extrapolations in $J$ are used for each temperature point. To this
end, the exactly known result of the free phonons is subtracted from 
the series obtained. The resulting expression  is extrapolated;
it starts in second order in $J$. Starting from results for the 
partition function in order 13 in $J$, the maximum order of extrapolation of
the remaining specific heat is 10. Ordinary Pad\'e extrapolations
would allow a maximum order of 11, but due to the differentiation for
the Dlog-Pad\'e extrapolations one additional order is lost. After the
extrapolation, the free phonon contribution is added again. This procedure
is chosen to deal with series which behave qualitatively like the
high-temperature series for pure spin systems.

In Fig.~\ref{fig:spec.Dlog} the results for three different sets of parameters
$\omega=1J$, $g=0.7$ (upper panels), $\omega=1J$, $g=1.5$ (middle
panels) and $\omega=5J$, $g=1$ (lower panels) are shown. The left plots depict 
the $[n,2]$ and the right plots the $[n,4]$ extrapolations. In the $[n,m]$ 
scheme the logarithmic derivative is approximated by a rational function where 
the polynomial in the numerator has the degree $n$ while the one in the 
denominator has the degree $m$.

The  $[n,2]$ and the $[n,4]$ scheme converge very well. The $[n,4]$ 
extrapolations are more stable in the low temperature regime for both 
parameter sets. Hence, this scheme is used in the following. The range
of validity can be specified by $T \gtrsim 0.15J$ as long as the
spin-phonon coupling is smaller or of the same order as $\omega$.  For
values of $gJ/\omega > 1$ the extrapolations suffer often from
spurious poles. In the middle panels the defective extrapolations are
visible. The schemes $[5,2]$, $[6,2]$, and $[3,4]$ yield extrapolations
which differ significantly from the other schemes considered. This is
due to spurious poles in the integration interval with respect to
$J$. For $[6,2]$ even temperatures above $T \approx J$ are not
described reliably.

For small spin-phonon coupling $g$ the specific heat is well
described by the sum of the phononic $C_B$ and the magnetic
 specific heats $C_S$ of the isolated subsystems, cf.\
first row of panels in  Fig.~\ref{fig:spec.Dlog}.
Increasing the spin-phonon coupling  shifts the specific heat to 
larger values, cf.\ second row of panels in  Fig.~\ref{fig:spec.Dlog}.
In this temperature regime ($T > J$) already the truncated series 
yield trustworthy results. 

Fixing the spin-phonon coupling $g$ and increasing
the phonon frequency $\omega$ weakens the visible effect of the
phonon dynamics, cf.\ third row of panels in  Fig.~\ref{fig:spec.Dlog}. 
This is to be expected since  the phononic subsystems becomes more rigid,
so that it will not be influenced significantly by the spin systems
and vice-versa.

\begin{figure}[htbp]
  \begin{center}
    \leavevmode
    \includegraphics[width=\columnwidth]{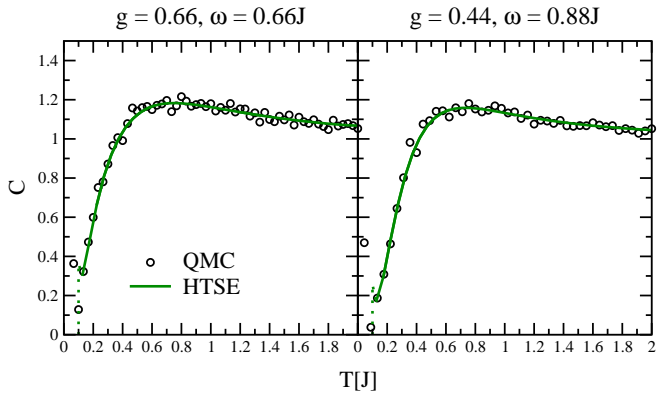}
    \caption{Dlog-Pad\'e extrapolations of the specific heat compared
      to QMC data: the left plot shows data for the values $g=0.66$,
      $\omega=0.66J$, and the right plot depicts the parameters
      $g=0.44$, $\omega = 0.88J$.}
    \label{fig:spec.cmp.QMC}
  \end{center}
\end{figure}
Finally, we compare the extrapolated results 
from the cluster expansion in $J$ to data obtained by QMC. 
Fig.~\ref{fig:spec.cmp.QMC}
depicts two sets of parameters of the $[6,4]$ scheme and the 
corresponding QMC data. The left panel shows the
result for $g=0.66$, $\omega=0.66J$. This parameter set implies
that the system is gapped \cite{raas02a}.

In the right panel results  are depicted
for $g=0.44$, $\omega=0.88J$ which implies a gapless phase.
The statistical error of the QMC data can be estimated from the spread
of the data points. The extrapolated series yield  smooth, continuous
results which agree very well with the QMC data. For temperatures
below $T/J \approx 0.15$ the dotted lines refer to defective
extrapolations which are depicted for illustration. These extrapolations
are known to yield unreliable results beforehand because the
extrapolants display spurious poles in the integration interval of $J$.
Note that each temperature requires an extrapolation. The extrapolations at
higher temperatures turn out to be very stable and not defective so that
reliable results can be obtained for temperatures $T \gtrsim 0.15 J$.  
For lower temperatures, the extrapolations are contaminated by
spurious poles and should not be trusted.

\subsection{\label{ssec:susc} Susceptibility}
The magnetic susceptibility $\chi$ is of special interest since it is
most easily accessible experimentally. Very often static models like the
dimerized and/or frustrated spin chain can already yield 
a good description of the susceptibility of one-dimensional systems.
Two cases are conceivable:
either the appropriate model is indeed a static spin model or
the static spin model should be seen as effective model which 
incorporates the effects of the spin-phonon coupling. 
Of course, it makes only sense to distinguish both cases if there
are other experimental probes to discriminate between them.
This will be elucidated in Sect.\ \ref{ssec:fit}.

In the antiadiabatic limit $\omega > J$, detailed investigations
of the susceptibility were done. The spin-phonon chain could be 
mapped to a frustrated spin chain with temperature dependent exchange
couplings \cite{raas02a}. The corresponding susceptibility could
be obtained from a high temperature series expansion \cite{buhle01a}.
It was shown that $\chi$
is only little affected by the temperature dependence of the coupling
constants. Thus it can be neglected and a static model is indeed well
justified. This finding agrees with previous results
\cite{uhrig01a}. It explains, for instance, why the magnetic susceptibility 
of CuGeO$_3$ can be fitted so surprisingly well
by a static frustrated Heisenberg model \cite{casti95,riera95,fabri98a}.

We expect that the mapping to a static spin model works less well
in the crossover regime to the adiabatic limit $\omega \lesssim J$.
There the effects of the phonon dynamics should be more clearly
visible. We will return to this question in Sect.\ \ref{ssec:fit}.

The series expansion of the susceptibility is obtained from the previous
considerations by incorporating an external magnetic field. 
In a first step, the coupling of this field to the spins is added to the
unperturbed Hamilton operator $H_0$ which leads to a modified
free energy series expansion. In a second step, the susceptibility can 
be derived from the free energy series. 

The unperturbed part
$H_0$ of the Hamilton operator (\ref{eqn:hamiltoninter}) is
extended by a magnetic field term leading to 
\begin{equation}
  \label{eqn:hamilton0chi}
  H_0 = \omega \sum_i b_i^\dagger b_i^{\phantom{\dagger}} -h\sum_iS_i^z
      = H_\text{B} -hM
\end{equation}
with the magnetic field $h$ given in units of $g\mu_B$. The additional
term proportional to the magnetization $M$ commutes with the
free phonon part $H_\text{B}$ and with the perturbation $V$ as
given in Eq.~(\ref{eqn:hamiltoninter}). Thus, the expression for
$\tilde{V}(\tau)$ in Eq.~(\ref{eqn:Vtilde}) is unchanged compared
to the calculation of the specific heat. The magnetic field is included in
$H_0$.

The formal expression Eq.~(\ref{eqn:partition}) for the
partition function is unchanged. From Eq.~(\ref{eqn:hamilton0chi})
we deduce the zeroth order contribution $Z_0$ which now
reads
\begin{subequations}
  \label{eqn:Z0chi}
  \begin{eqnarray}
    Z_0 &=& \text{tr}\left( e^{-\beta H_0}\right) = \text{tr}\left(
      e^{-\beta H_\text{B}} e^{\beta h M}\right) \\
    &=&
    z_0^N\left(2\text{cosh}\left(\frac{\beta h}{2}\right)\right)^N 
    \ .
  \end{eqnarray}
\end{subequations}
Taking the logarithm of the partition function $Z$ yields
\begin{subequations}
  \label{eqn:lnZchi}
  \begin{eqnarray}
    &&\frac{1}{N}\text{ln}Z = \text{ln}z_0 +
    \text{ln}\left(2\text{cosh}\left( \frac{\beta h}{2}\right)\right)
    \\
    &&+\sum_{n=1}^\infty\left(-J\right)^n \left\{
      \int\limits_0^{\beta}\!d\!\tau_1\cdots
      \int\limits_0^{\tau_{n-1}}\!\!d\!\tau_n \langle\tilde{V}(\tau_1)
      \cdots \tilde{V}(\tau_n) \rangle \right\} \hspace*{8mm}
  \end{eqnarray}
\end{subequations}
with $z_0$ as given in Eq.~(\ref{eqn:partition0}). The angular
brackets $\langle \cdots \rangle$ denote the coefficients proportional
to $N$ in the trace, see Eq.~(\ref{eqn:Vbrackets}). To derive the
susceptibility the above equation has to be differentiated twice
with respect to the magnetic field $h$. Finally $h$ is set to zero
\begin{equation}
  \label{eqn:chi}
  T\chi = \frac{1}{\beta^2}\frac{\partial^2}{\partial h^2} \left(
  \frac{1}{N}\text{ln}Z\right)\Bigg|_{h=0} \ .
\end{equation}

The susceptibility could be computed up to order 12 in $J$. The
contribution of 2242 connected and of 2810 disconnected clusters had to
be evaluated. To illustrate the result the first orders of the susceptibility
series are given
\begin{eqnarray}
  \nonumber
  && T\chi = \frac{1}{4} -\frac{1}{8}J\beta 
    - \frac{1}{16}J^2 \beta \frac{g^2 }{\omega}
    + \frac{1}{96}J^3\beta^3
    \\
   &&\quad + J^4\frac{1}{1536} \Bigg( 5\beta^4+24\frac{g^2\beta^3}{\omega} 
   \nonumber \\
    &&\quad + \Big( \left( -72 {z_0^2} + 72{z_0} \right) 
      \frac{g^4}{\omega^2}
   + \left( -96 + 192{z_0} \right) \frac{g^2}{\omega^2} \Big)
   \beta^2 \nonumber \\
   &&\quad + \left(\left( 72{z_0} - 36
    \right) \frac{g^4}{\omega^{3}} - 192\frac {g^2}{\omega^3} \right)
    \beta  \Bigg) + \mathcal{O}\left(J^5\right)\ .
    \label{eqn:chifirst}
\end{eqnarray}

\begin{figure}[htbp]
  \begin{center}
    \leavevmode
    \includegraphics[width=\columnwidth]{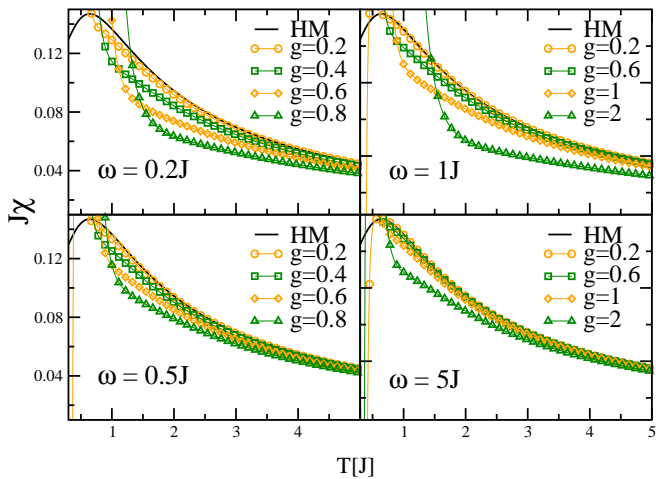}
    \caption{Truncated series of the susceptibility: various sets of
      parameters are shown. The exactly known result for the
      Heisenberg model serves as a reference. The left panels
      illustrate the adiabatic limit whereas the right panels show the
      results in the antiadiabatic limit.} 
    \label{fig:chiseries}
  \end{center}
\end{figure}
In Fig.~\ref{fig:chiseries}, the truncated susceptibility series
are depicted for various parameter sets. The energy scales are given in
units of the magnetic exchange coupling $J$.  The exact result of the
Heisenberg model serves as a reference to illustrate the effects of
the additional coupling to the phonons. The general feature that the
results diverge for temperatures below $T\lesssim 1.5J$ is expected
for the truncated series. But the qualitative behavior of the
susceptibility is already discernible.

The left panels depict the adiabatic regime and the right panels
illustrate the antiadiabatic limit. The following conclusions can
be drawn from the truncated series  in the temperature regime $T\gtrsim 1.5J$.
Fixing the phonon frequency $\omega$ the
overall height of the susceptibility is lowered for increasing
spin-phonon coupling $g$. Such a behavior can be understood in a mean-field
treatment of the spin-phonon coupling. For increasing $g$ the
effective coupling $J_\text{eff} = J(1+g\langle b^\dagger
+ b^{\phantom{\dagger}}\rangle)$ increases. This shifts the whole
susceptibility to lower temperatures compared to the result of the
Heisenberg model with the bare magnetic exchange $J$,
 cf.\ discussion in Ref.~\cite{raas02a}. 

For fixed spin-phonon coupling and
increasing phonon frequency $\omega$ this effect becomes less
pronounced. For increasing phonon frequency the magnetic and phononic
degrees of freedom decouple more and more because the phonon system 
becomes increasingly rigid so that it is  influenced  less and less  by
the magnetic subsystem. Concomitantly, the spin system is less influenced
by the phononic subsystem. Thus, for fixed $g$ and $\omega \rightarrow \infty$ 
the magnetic properties are  dominated by the 
antiferromagnetic Heisenberg model.  

\begin{figure}[htbp]
  \begin{center}
    \leavevmode
    \includegraphics[width=\columnwidth]{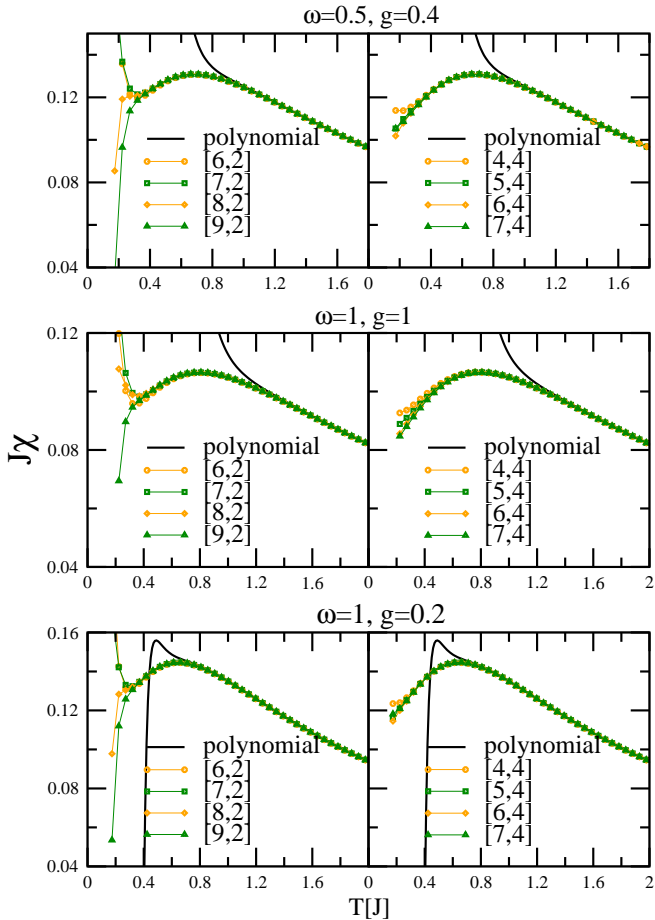}
    \caption{Dlog-Pad\'e extrapolations of the susceptibility: various
      extrapolation schemes are compared for three different sets of
      parameters. The  parameters in the first two rows of panels
      correspond to the gapped regime whereas the lowest row depicts
      results in the gapless regime.}
    \label{fig:susc.Dlog}
  \end{center}
\end{figure}
The truncated series fail to reproduce the very significant maximum
of $\chi(T)$, cf.\ Fig.~\ref{fig:chiseries}.
 This maximum serves as a landmark for many experimental
analyses. Thus the height and the position of the
maximum are of great interest. Extrapolations are necessary to 
extend the representation of the results beyond the radius of
convergence of the truncated series so that the maximum can be
captured reliably. For the susceptibility the same
extrapolation techniques are applied as for the pure spin models
described in Ref.~\cite{buhle01a}. Basically, the truncated series
is extrapolated using Dlog-Pad\'e approximants in an Euler-transformed
variable. In contrast to the pure spin models the extrapolations of
the results of the spin-phonon model is not performed in the
inverse temperature $\beta$, but in the magnetic exchange coupling
$J$. For each temperature a separate
extrapolation in $J$ has to be done. Using standard routines from
computer algebra programs this does not pose more problems than the
previous extrapolations in $\beta$.

A more serious restriction concerns the behavior at large values of
$J$. For the pure spin systems it was very efficient
to bias the extrapolations in the inverse temperature such that
the known low temperature behavior was captured, see e.g.\
Ref.~\cite{buhle01a}. For the spin-phonon system, however, much
less is known about the excitations at low-energies. From
the phase diagram  depicted in Ref.~\cite{raas02a} it can be deduced
whether the system is gapped or gapless. But the precise value of
the gap, let alone the form of the dispersion, are not available
in the limit $J\to\infty$. Note that for the present expansion it is
this limit that we need to understand, not the limit $T\to0$ as in
the high-temperature series.
Hence, we do not attempt to bias our expansions in their behavior
at large values of $J$. But an improved understanding of the 
limit $J\to\infty$ will certainly help to obtain even better
extrapolations.
 
Fig.~\ref{fig:susc.Dlog} shows an overview over the susceptibilities 
obtained  from unbiased Dlog-Pad\'e extrapolations for three different 
sets of parameters. The upper two rows correspond to results
in the gapped regime whereas the parameters in the last row
correspond to the gapless regime, see the phase diagram in
Ref.~\cite{raas02a}. The left panels depict the extrapolations of
the form $[n,2]$ and the right panels the extrapolations of the form
$[n,4]$. 

The position and the height of the maximum are described reliably by both 
extrapolation schemes $[n,2]$ and $[n,4]$. This conclusion is based on the 
agreement of the results for different orders $n$ in Fig.\ \ref{fig:susc.Dlog}.
We observe that the $[n,4]$ extrapolations converge better than the 
$[n,2]$ extrapolations for increasing order. 
Higher values of $m$ in the general $[n,m]$ scheme  or odd values of $m$
are likely to imply spurious poles. Thus, the $[n,4]$ scheme is
used in the following to represent the susceptibility for all parameter
sets shown in this paper. 

For temperatures $T<0.2J$  no results are depicted due to spurious poles in the
extrapolations in $J$. The range of validity of the $[n,4]$ extrapolations can 
be estimated to be at least $T\gtrsim 0.25J$ independent of the parameter sets 
that we analyzed in our investigations.

\begin{figure}[htbp]
  \begin{center}
    \leavevmode
    \includegraphics[width=\columnwidth]{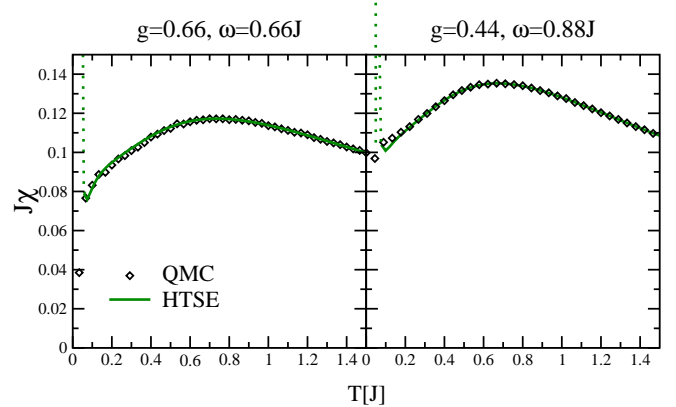}
    \caption{Dlog-Pad\'e extrapolations of the susceptibility compared
      to QMC data: the left plot shows data for the values $g=0.66$,
      $\omega=0.66J$ (gapped phase), and the right plot depicts data for the
      values $g=0.44$, $\omega = 0.88J$ (gapless phase).}
    \label{fig:susc.cmp.QMC}
  \end{center}
\end{figure}
Finally, we illustrate the reliablity of our extrapolations by a comparison
to QMC data. For two different sets of parameters Fig.~\ref{fig:susc.cmp.QMC} 
displays the $[7,4]$ extrapolations of the series and the corresponding QMC 
data. The values $g=0.66$, $\omega =0.66J$ in the left panel imply that the 
system is gapped. The values $g=0.44$, $\omega =0.88J$ in the right
panel correspond to the gapless regime. The agreement between 
the extrapolated series results and the QMC data is very good. 
For temperatures above $T/J \gtrsim 0.2$ the results coincide. 
Below $T \lesssim 0.2 J$ the QMC data  and the series data deviate from each 
other. At present, we cannot decide whether the deviations for $0.1J < T <0.2J$
are due to problems in the QMC simulation like statistical errors 
or finite-size effects or whether they are due to problems in the series 
extrapolations. Below $T\approx 0.1J$, spurious poles occur in the
extrapolated integrands leading to defective extrapolations. 
For illustration, these defective extrapolations are shown as dotted lines
in Fig.~\ref{fig:susc.cmp.QMC}.

We emphasize that the convincing agreement of the QMC and the series 
results in the regime  $T \gtrsim 0.2 J$ supports the reliability
of these results. In particular, the position and the height of
the important maximum of the susceptibility is  described quantitatively.

\subsection{\label{ssec:fit} Comparison to Static Spin Models}
In this paragraph the importance of the spin-phonon dynamics shall be
highlighted. As mentioned before it is often possible to describe
the properties of a spin-phonon model by a static spin model alone,
in particular in the antiadiabatic regime. It is to be expected that
the effects of the dynamic nature of the spin-phonon coupling are
most prominent in the regime where all energies are of similar magnitude.
To provide evidence for this hypothesis we perform the following
``theoretical'' experiment.

\begin{figure}[htbp]
  \begin{center}
    \leavevmode
    \includegraphics[width=0.95\columnwidth]{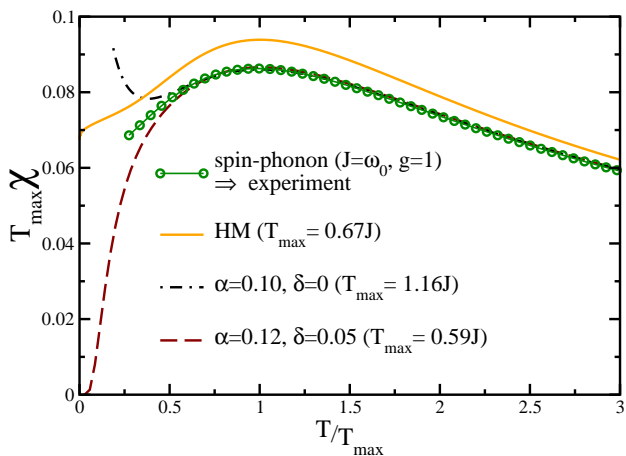}
    \caption{``Theoretical'' experiment: The data
      for $g=1$ and $\omega_0 = J$ (symbols) is taken as 
      (mock) experimental input. It is fitted by the
      susceptibility of static spin model (solid lines). The temperature
      $T_\text{max}$ denotes at which temperature the susceptibility has
      its maximum; this value sets the natural energy scale of the problem.}
    \label{fig:susc.fit}
  \end{center}
\end{figure}

We start from the extrapolated series data for $g=1$ and $\omega_0 = J$,
see symbols in Fig.\ \ref{fig:susc.fit}.
These data shall serve as ``experimental'' input which we will then
analyze using static spin models. The static spin model considered here
is the frustrated and dimerized spin chain given by
\begin{equation}
H = J\sum_i\left[ (1 + \delta (-1)^i)\mathbf{S}_{i}\mathbf{S}_{i+1} +
     \alpha \mathbf{S}_{i}\mathbf{S}_{i+2}\right] \ .
\end{equation}
In this way, we imitate the
standard procedure one would apply to experimental data. The objective
is to see to which extent such an analysis yields agreement.
In particular, we are interested to see where such a description
remains unsatisfactory. Such an unsatisfactory description based on
static spin models is the signature of the dynamic nature of the spin-phonon
coupling.

In Fig.~\ref{fig:susc.fit} the ``experimental'' susceptibility is analyzed
by the susceptiblity of static spin models, i.e.\ the isotropic Heisenberg 
model (HM), the purely frustrated spin chain and the dimerized and
frustrated chain. Obviously, the HM is not appropriate to
describe the data depicted by the symbols. But the other two parameter sets
($\alpha=0.1,\delta=0$) and ($\alpha=0.12,\delta=0.05$) describe the
experimental data very well for not too low temperatures, i.e.\ 
$T>0.5T_{\text{max}}$. This temperature regime corresponds to the
the range of temperatures where the extrapolated high-temperature
series  for the dimerized and frustrated spin chain are reliable
\cite{buhle01a}. 

Of course, it is possible to distinguish between different models if 
reliable (experimental) data down to low temperatures are available.
In practice, however, this is often not the case since the data at
low temperatures can be contaminated by impurity effects or other
imperfections like inclusions of other phases or simply the presence
of other structural elements in the sample. In such a situation, the
determination of the appropriate microscopic model is difficult and
ambiguities are hard to avoid.

\begin{figure}[htbp]
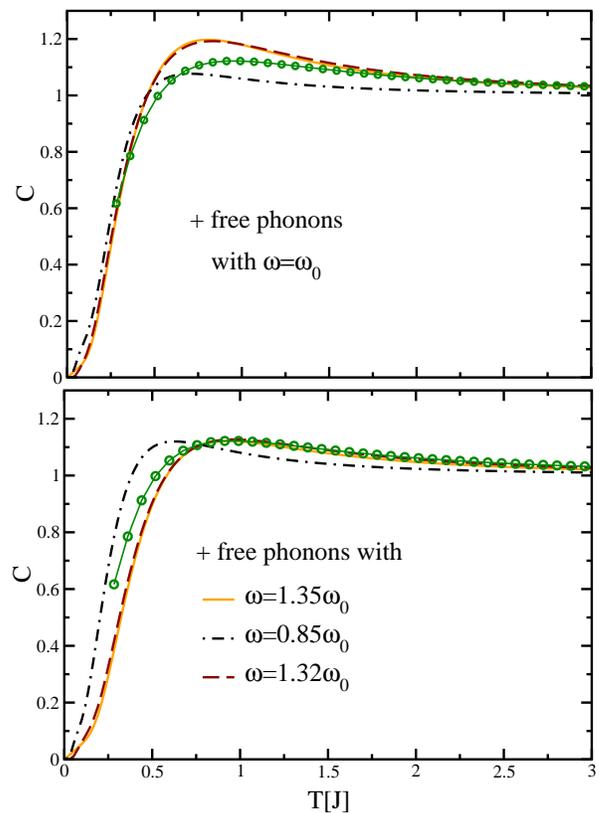

  \begin{center}
    \leavevmode
    \includegraphics[width=0.89\columnwidth]{./fig_buhler8a.eps}\\
    \includegraphics[width=0.9\columnwidth]{./fig_buhler8b.eps}
    \caption{Comparison of the specific heat for the parameter sets
      determined from the susceptibility in Fig.\ \ref{fig:spec.susc.fit}.
      For the upper plot the phonon frequency is assumed to  be known
      beforehand. For the lower plot no a priori knowledge is
      assumed for the phonon frequency; so it is also fitted. 
      The line styles are the same as those used in Fig.~\ref{fig:susc.fit},
      that means the exchange couplings used are those from the corresponding
      curves in Fig.~\ref{fig:susc.fit}.} 
    \label{fig:spec.susc.fit}
  \end{center}
\end{figure}

One way to make progress is to use the parameter set determined from the
susceptibility data and to examine whether other properties can be
understood with the same parameter set as well. Here we choose to 
study the specific heat.  In the upper panel of
Fig.~\ref{fig:spec.susc.fit} the phonon frequencies are assumed to be
known a priori; so the free phonon contribution added
is the one for this known frequency $\omega=\omega_0$.
Clearly, none of the static spin models describes the
susceptibility data \emph{and} the specific data satisfactorily.
From the knowledge of both quantities compelling evidence can
be deduced that a dynamic spin-phonon coupling must be present.

But the situation can be less advantageous in practice. Let us assume
that we do not possess knowledge about the frequency of the phonon
to which the spin system may or may not be coupled. Then this frequency
could be seen as an additional fit parameter. This view point was adopted
in the lower plot in Fig.~\ref{fig:spec.susc.fit}. The parameter set
($\alpha=0.1; \delta=0$) of the dashed-dotted curve can be clearly discarded.
The parameter set ($\alpha=0; \delta=0$, the Heisenberg model HM) 
of the solid curve can be discarded because it does not describe the 
susceptibility, cf.\ Fig.~\ref{fig:susc.fit}. The parameter set 
($\alpha=0.12; \delta=0.05$) of the dashed curve seems to fit the data
down to $T\approx 0.7J$. For lower values, however, the agreement is poor.
So also this data set must be discarded. Note that we refer only to the
temperature regime where the  extrapolated series yield reliable results.

So we have shown for the above example that it is \emph{not} possible
to describe the data of a dynamic spin-phonon system in the intermediate
regime, where all energy scales are of similar magnitude,
in the framework of static spin models plus (decoupled) phonons.
But it is not sufficient to consider only one quantity at (relatively)
high temperatures. In order to obtain unambiguous evidence of the
presence of a dynamic spin-phonon coupling one has to dispose of
either data down to low temperatures (not considered here) 
or to consider at least two
independent thermodynamic quantities. Otherwise, one may easily
be misled by a good agreement in one quantity alone to conclude
that a simpler static model is sufficient for a microscopic description.

\section{\label{sec:conclusion} Summary and Outlook}
The problem of a one-dimensional spin system coupled to phononic degrees 
of freedom is investigated. A cluster expansion is applied to obtain
a series expansion in the magnetic coupling $J$. Results are computed
at finite temperatures for the free energy, the specific heat, and 
magnetic susceptibility. No truncation in the phononic
subspace is necessary since the expansion is performed in
$J$ about the limit $J=0$, not in the inverse temperature.
This is the first realization of a cluster expansion at finite 
temperatures for an extended spin-phonon
problem. The implementation of the expansion in a computer 
program yields high orders in the expansion parameter. 

The comparison of the results obtained from the extrapolated
series  to data from quantum Monte-Carlo simulations shows a very 
good agreement. This supports the reliability of the approach. 
Hence our results can serve as input for quantitative data analysis
since the features of the magnetic susceptibility and the specific
heat at moderate and at high temperatures $T\gtrsim 0.15 -0.25J$
are described reliably. 

A possible route to improve the extrapolations is to understand
the limit $J\to\infty$ in the Hamilton operator (\ref{eqn:hamilton})
better. At first glance, this limit looks simple since it corresponds
to the adiabatic situation with $\omega/J \to 0$. But this limit is
not straightforward since the nature of the excitations is unclear
at present. Note that in this limit static displacements can be
made at no energetic cost. This suggests that the excitations
are domain walls. Whether these objects are static (because
$\omega/J\to 0$) or dynamic (because the magnetic subsystems
retains its fluctuations) is unclear at present and constitutes
an interesting theoretical issue in itself.

Finally, we analyzed the data of a spin-phonon system in the intermediate
coupling regime, where all energies are of similar magnitude, in great detail. 
As expected, we could show that the effects of the dynamic spin-phonon
coupling cannot be imitated by a static spin model and decoupled 
phonons. But it is necessary to study low temperatures or at least
two quantities like the susceptibility and the specific heat carefully.
Otherwise, one may be easily misled by a good agreement in one
quantity alone to conclude that a static microscopic model is sufficient.
We think that this conclusion is helpful for future experimental
analyses of low-dimensional spin systems.

\begin{acknowledgments}
We like to thank C.\ Aits and U.\ L\"ow for kindly providing the
QMC data as benchmarks. We acknowledge the financial support by
the DAAD for an extended visit by one of us (AB)
to the University of New South Wales where a significant part of this project 
has been carried out. We are indebted to the DFG for the support
in SP1073 and SFB608. Last, but not least, one of us (GSU) 
gratefully acknowledge the hospitality and the support of the COE at 
Tohoku University, Sendai, where the manuscript has been finished.
\end{acknowledgments}


\end{document}